\documentclass[traditabstract, longauth]{aa} % for the abstract without structuration 
\usepackage{txfonts}
\usepackage{graphicx}
\usepackage{caption}
\usepackage{subcaption}
\usepackage{color}
\usepackage{natbib}
\bibpunct{(}{)}{;}{a}{}{,}
\def\gapprox{{_>\atop{^\sim}}}
\def\lapprox{{_<\atop{^\sim}}}

\def\cmmd{\rm {cm^{-3}}}
\def\cmmt{\rm {cm^{-2}}}

\def\s-1{\rm {s^{-1}}}

\def\HC3N{HC$_3$N}

\def\kms{\hbox{${\rm km\,s}^{-1}$}}
\def\msun{M$_{\odot}$}
\def\lsun{L$_{\odot}$}

\newcommand{\asec}{\mbox{$''$}}

\usepackage{url}

\begin{document}
 \title{Luminous, pc-scale CO 6--5 emission in the obscured nucleus of NGC~1377}
%\thanks{Based on observations
%carried out with the ALMA Interferometer. ALMA is a partnership of ESO (representing its member states),
%NSF (USA) and NINS (Japan), together with NRC (Canada) and NSC and ASIAA (Taiwan),
%in cooperation with the Republic of Chile. The Joint ALMA Observatory is operated by ESO, AUI/NRAO and NAOJ.}}

\author{S. Aalto
          \inst{1}
          \and
        S. Muller\inst{1}
          \and
	  F. Costagliola\inst{1}
          \and
          K. Sakamoto\inst{2}
	 \and
          J. S. Gallagher\inst{3}
        \and
	N. Falstad\inst{1}
	\and
	S. K\"onig\inst{1}
	\and
        K. Dasyra\inst{4}
       \and
	K. Wada\inst{5}
	\and
         F. Combes\inst{6}
         \and
	  S. Garc\'ia-Burillo\inst{7}
        \and
         L. E. Kristensen\inst{8}
	\and
	 S. Mart\'in\inst{9,10,11}
        \and	
          P. van der Werf\inst{12}
	\and
 	A. S. Evans\inst{13}
	\and
        J. Kotilainen\inst{14}
             }

 \institute{Department of Earth and Space Sciences, Chalmers University of Technology, Onsala Observatory,
              SE-439 92 Onsala, Sweden\\
              \email{saalto@chalmers.se}
\and  Institute of Astronomy and Astrophysics, Academia Sinica, PO Box 23-141, 10617 Taipei, Taiwan 
\and  Department of Astronomy, University of Wisconsin-Madison, 5534 Sterling, 475 North Charter Street, Madison WI 53706, USA
\and Department of Astrophysics, Astronomy \& Mechanics, Faculty of Physics, University of Athens, Panepistimiopolis Zografos 15784, Greece
\and Kagoshima University, Kagoshima 890-0065, Japan
\and Observatoire de Paris, LERMA (CNRS:UMR8112), 61 Av. de l'Observatoire, 75014 Paris, France 
\and Observatorio Astron\'omico Nacional (OAN)-Observatorio de Madrid, Alfonso XII 3, 28014-Madrid, Spain
\and Centre for Star and Planet Formation, Niels Bohr Institute and Natural History Museum of Denmark, University of Copenhagen, {\O}ster Voldgade 5-7, DK-1350 Copenhagen K, Denmark
\and European Southern Observatory, Alonso de Córdova 3107, Vitacura, Santiago, Chile
\and Joint ALMA Observatory, Alonso de Córdova 3107, Vitacura, Santiago, Chile
\and Institut de Radio Astronomie Millim\'etrique (IRAM), 300 rue de la Piscine, Domaine Universitaire de Grenoble,
38406 St. Martin d$ ' $H\`eres, France
\and Leiden Observatory, Leiden University, 2300 RA, Leiden, The Netherlands
\and University of Virginia, Charlottesville, VA 22904, USA, NRAO, 520 Edgemont Road, Charlottesville, VA 22903, USA
\and Finnish Centre for Astronomy with ESO (FINCA), University of Turku, V\"ais\"al\"antie 20, FI-21500 Kaarina, Finland
 }

   \date{Received xx; accepted xx}

  \abstract{High resolution submm observations are important in probing the morphology, column density and dynamics of obscured active galactic nuclei (AGNs). With high resolution
($0.\asec 06 \times 0.\asec 05$) ALMA 690~GHz observations we have found bright ($T_{\rm B}>$80 K) and compact 
(FWHM 10$\times$7 pc) CO 6--5 line emission in the nucleus of the extremely radio-quiet galaxy NGC~1377.  The CO 6--5 integrated intensity is aligned with the previously discovered
jet/outflow of NGC~1377 and is tracing the dense ($n>10^4$ $\cmmd$), hot gas at the base of the outflow. The velocity structure is complex and shifts across the jet/outflow are discussed
in terms of jet-rotation or separate, overlapping kinematical components.
High velocity gas ($\Delta v \pm 145$ \kms) is detected inside $r<$ 2-3 pc and we suggest that it is emerging from an inclined rotating disk or torus
of position angle PA=140$^{\circ} \pm 20^{\circ}$ with a dynamical mass of $3 \times 10^6$ \msun. This mass is consistent with that of a supermassive black hole (SMBH), as inferred from
the $M-\sigma$ relation. The gas mass of the proposed disk/torus constitutes $<$3\% of the nuclear dynamical mass.
In contrast to the intense CO 6--5 line emission, we do not detect dust continuum with an upper limit of $S$(690~GHz)$\lapprox$2mJy. The corresponding, 5 pc, H$_2$ column
density is estimated to $N$(H$_2$)$<3 \times 10^{23}$ $\cmmt$, which is inconsistent with a Compton Thick (CT) source. We discuss the possibility that CT obscuration may be occuring
on small (subparsec) or larger scales.
From SED fitting we suggest that half of the IR emission of NGC~1377 is nuclear and the rest (mostly the far-infrared (FIR)) is emerging from larger scales.  The extreme radio quietness, and
the lack of  emission from other star formation tracers, raise questions on the origin of the FIR emission. We discuss the possibility that it is arising from the dissipation of shocks in the molecular jet/outflow or
from irradiation by the nuclear source along the poles. 
 }


 \maketitle

%________________________________________________________________

\section{Introduction}
\label{s:intro}

NGC~1377 is a nearby (21~Mpc (1\arcsec=102~pc)), lenticular galaxy with a far-infrared (FIR) luminosity of $L_{\rm FIR}=4.3 \times 10^9$ L$_{\sun}$ \citep{roussel03}.
NGC~1377 is  the most radio-quiet, FIR-excess galaxy known to date with radio emission being deficient by a factor $\approx$37 with 
respect to normal galaxies \citep{roussel03,roussel06}.
Its nucleus is dust enshrouded \citep[e.g.][]{spoon07} and the source of the FIR luminosity (and cause of its radio deficiency) has remained elusive. A nascent starburst \citep{roussel03,roussel06}
or a radio-quiet AGN \citep{imanishi06,imanishi09} have been proposed as possible solutions.  The energetics of a powerful molecular outflow seem to point towards an AGN
\citep{aalto12b} and recent ALMA observations revealed that at least part of the outflow is in the form of a peculiar molecular jet. Velocity reversals along the jet
may be indications of precession \citep{aalto16}. Recently, faint radio emission with a synchrotron spectrum has been detected for the first time, but there appears to be no nuclear
X-ray source \citep{costagliola16}.  The radio detection confirms the extreme FIR-excess of NGC~1377 and that it is well away from the radio-FIR correlation \citep{helou85}. 
\citet{costagliola16} use the radio emission to estimate the star formation rate (SFR) finding an SFR$<$0.1 \msun\ yr$^{-1}$, which is not sufficient to power the observed IR luminosity and
to drive the CO outflow. However, even if the evidence in support of a buried AGN is mounting, the nature of the power source behind the FIR luminosity and
the molecular outflow is still not fully understood. 
%Why is the nucleus faint in both radio- and X-ray emission?

Observations at mid-infrared (mid-IR) wavelengths reveal a compact ($<$0.\asec 14), high surface brightness source \citep{imanishi11} in the nucleus of NGC~1377. It is not clear if this structure is part
of an obscuring torus or disk and if it is opaque enough to absorb X-rays emerging from an accreting supermassive black hole. It is also not understood if and how the obscuring material
is linked to the outflowing gas in the jet. 

Submm observations of dust continuum are powerful tools in detecting the presence of opaque structures of obscuring dust \citep[e.g.][]{sakamoto08,wilson14}. Direct detection on
pc-scales of the dust torus around an AGN have, however, remained elusive until recently when Atacama Large Millimeter/submillimeter Array (ALMA) band 9 (690~GHz) observations
%of the Seyfert galaxy
of the Seyfert galaxy NGC~1068 revealed the presence of the obscuring torus and its turbulent dynamics \citep{garcia16, gallimore16}. We obtained  ALMA  band 9 observations to probe
the structure and orientation of the intervening dust in the nucleus of NGC~1377 and to determine gas and dust column densities. We also acquired simultaneous CO 6--5 observations to
probe the very nuclear pc-scale dynamics of the hot gas.

In Sections~\ref{s:obs} and \ref{s:res} we present the observations and the results. In Section~\ref{s:nucleus} we discuss the nuclear gas excitation, the nuclear dust obscuration and limits to the
associated H$_2$ column densities. We also present a simple two-component model to the dust Spectral Energy Distribution (SED) and discuss the origin of the far-infrared (FIR) emission of NGC~1377.
In Section~\ref{s:hivel} we examine the possible source of the CO 6--5 high-velocity gas, the dynamics of the nuclear disk and the accleration region of the jet/outflow.

%__________________________________________________________________

\section{Observations}

\label{s:obs}

Observations of the CO J=6--5 line were carried out with ALMA (with 35 antennas in the array) on 2015 September 25, for
4 minutes on-source (30 minutes in total) and with reasonable atmospheric conditions
(precipitable amount of water vapour of $\sim$0.5~mm, $T_{\rm sys}$=600-1200 K). The phase centre was set to $\alpha$=03:36:39.074 
and $\delta$=$-$20:54:07.055 (J2000).

The correlator was set up to cover two spectral windows of 1.875~GHz in spectral mode, one centred at a
frequency of $\sim$690~GHz to cover the CO $J$=6--5 line (in the lower side band), and the other
centred at 709~GHz to cover the HCN 8--7 line. In addition, two 2~GHz bands were set up in continuum mode.
%i.e., with a coarser velocity resolution.

The bandpass of the individual antennas was derived from the quasar $J0522-3627$. The quasar
$J0348-2749$ was observed for complex gain calibration.
The absolute flux scale was calibrated using the quasar $J0334-401$.  
%The flux density for $J0334-401$ was extracted from the ALMA flux-calibrator database.

After calibration within the CASA reduction package \citep{mcm07}, the visibility set was imported into the AIPS
package for further imaging. The synthesized beam is $0.\asec 06 \times 0.\asec 05$ (6$\times$5 pc)
with Briggs weighting (parameter robust set to 0.5). The on-source observation was short (250 s) - designed to detect a
purported high-surface brightness dusty nucleus. The large number of telescopes in the observation ensures a good
sampling of the uv-plane despite the short observation (B$_{\rm max}$=2250 m, B$_{\rm min}$= 41 m). However,  the
recoverable scale is still limited to theoretically 2.\asec 2, but effectively we could not recover structures larger than 1\asec.

The resulting data would have a maximum theoretical sensitivity of 9~mJy (dual polarization, Briggs weighting) per beam in a 20~\kms\ (47~MHz) channel width and in the highest grade
weather. Our data set is close to this with an rms of 10 mJy. This means that we are only sensitive to structures with high surface brightness (10 mJy =$T_{\rm B}$=10 K).

%-----------------------------------------------Results------------------------------------------------------------------------

%---------------------------------------------------------------------------------------------------

\begin{figure}[tbh]
\includegraphics[scale=0.72]{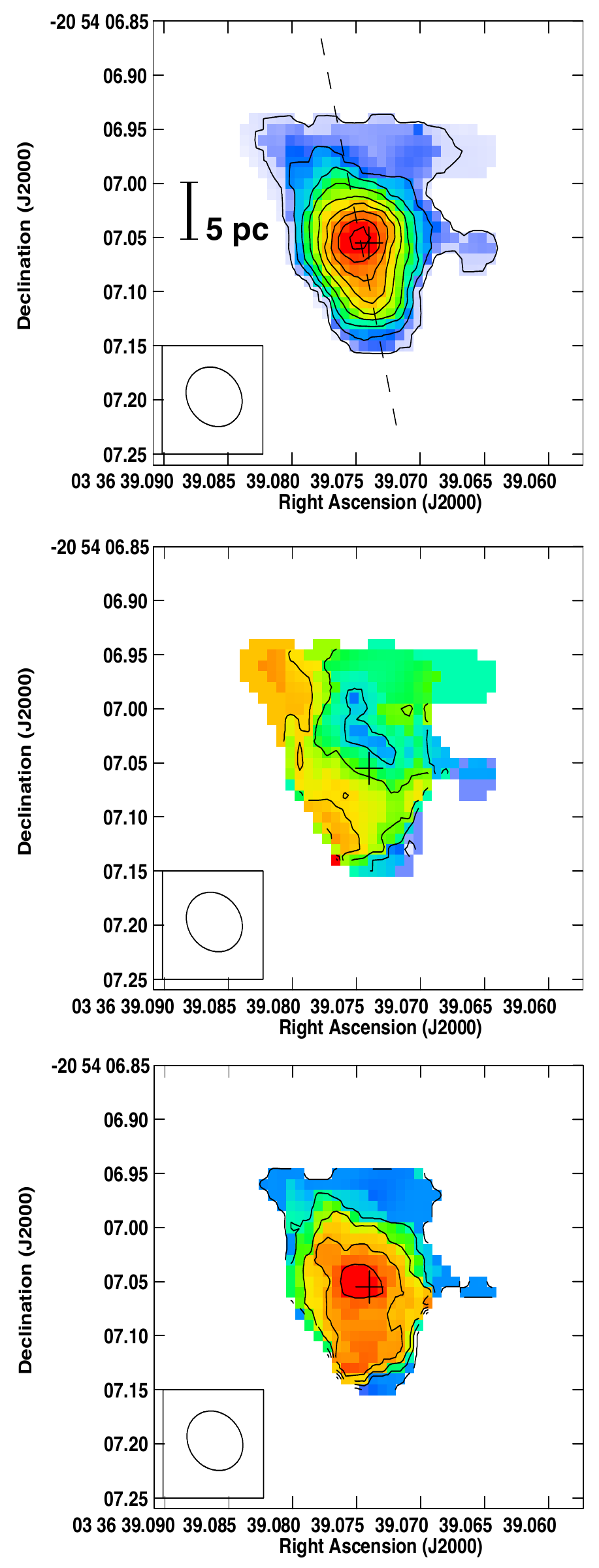}
\caption{\label{f:mom} \footnotesize CO 6--5 moment maps. Top: Integrated intensity (mom0) where contours are 0.6$\times$ (1,4,7,10,13,16,19,22,25)  Jy \kms\ beam$^{-1}$. Colours range
from 0 to 15 Jy \kms\ beam$^{-1}$. The dashed line marks the orientation of the molecular jet (see text for details). Centre: velocity field (mom1) where contours range from 
1690 \kms\ to 1820 \kms\ in steps of 10 \kms. Bottom: Map of the one-dimensional velocity dispersion (mom2), i.e. the FWHM line width of the spectrum divided by 2.35 for a Gaussian line profile. 
Contours are  4.4$\times$(1, 3, 5, 7, 9, 11, 13) \kms\ and colours range from 0 to 66 \kms. The cross indicates the position of the 345 GHz continuum peak (see Table~\ref{t:flux}).
}
\end{figure}

\section{Results}
\label{s:res}

\begin{table}
\caption{\label{t:flux} {CO 6--5 flux densities$^a$}}
\begin{tabular}{ll}
% & \\
\hline
\hline \\ 
Position (J2000) & $\alpha$:  03:36:39.075 ($\pm$ 0.\arcsec 01) \\
                           & $\delta$: -20:54:07.06 ($\pm$ 0.\arcsec 01) \\
Peak flux density & $90\pm 17$  mJy\,beam$^{-1}$\\ 
Flux &  \\
\, \, (central beam) & $15 \pm 2.5 $ Jy \kms\ beam$^{-1}$\\
\, \, (whole map) & $40 \pm 7.5$ Jy \kms\\

%\\

\hline \\
\end{tabular} 

\footnotesize {\it a)}\,  The position refers to the peak of the integrated CO 6--5 line emission. The Jy to K conversion in the $0.\asec 06 \times 0.\asec 05$ beam
is 1~K=1.1~mJy. The peak $T_{\rm B}$ is 80  K corresponding to 90 mJy. 
%We estimate the calibration errors to 20\%.

\end{table}

\subsection{CO 6--5}

We detect luminous CO 6--5 emission inside $r$=10 pc of the NGC~1377 nucleus. The peak flux density is $90\pm 17$  mJy\,beam$^{-1}$, which
corresponds to a brightness temperature $T_{\rm B} > 80$ K (Tab.~\ref{t:flux})\footnote{There are significant effects of decorrelation in our CO 6--5 data
implying that the line intensity (and hence $T_{\rm B}$) may be higher.  Here, we use 80 K as a lower limit for the CO 6--5 $T_{\rm B}$ in the 5 pc beam.}.

%-----
\subsubsection{Moment maps }
%integrated from 1500 to 2000 km/s (ch 9-33), cell 2 2, 3sigma: 4.7 mJy, file: 30, levs 0.1 (1,4,7,10,13,16,19,22,25) Jy km/s (in uncorrected scale)
%colour 0 - 2.5e3 (uncorrected scale)
%blc 232,236,trc 277, 275

The CO 6--5 integrated intensity (moment 0) map, velocity field (moment 1) and dispersion map (moment 2) are presented in Fig~\ref{f:mom}.
We clipped the moment 0 map at the 3$\sigma$ level. The velocity centroids were determined through a flux-weighted
first moment of the spectrum of each pixel, therefore assigning one velocity to a spectral structure. The dispersion was determined through a
flux-weighted second moment of the spectrum of each pixel.
CO 6--5 flux densities are presented in Tab.~\ref{t:flux}. 

{\it Moment 0 map} \, We find a centrally peaked structure with extensions to the north-east and south-west roughly consistent with the orientation of the previously found
molecular jet \citep{aalto16}. There is also a feature to the north-east with a position angle (PA) of 30$^{\circ}$ and fainter emission fanning out to the north-west 
of the nucleus. With a two-dimensional Gaussian fitting we find a full width half maximum (FWHM) source size of  0.\asec096 $\times$ 0.\asec 07 ($\pm 0.01$) (10 $\times$ 7 pc). The
position angle is PA=20$^{\circ} \pm15 ^{\circ}$. 
%

%
%integrated from 1500 to 2000 km/s (ch 9-33), cell 2 2, 3sigma: 4.7 mJy, file: 30, levs 1.71, 1.73, 1.75, 1.77, 1.79, 1.81x 1000e3 km/s colour 1700e3-1815e3,
% km/s, blc 232,236,trc 277, 275
%

{\it Moment 1 and 2 maps} \, The centroids of the velocity field span 1690 - 1820 \kms.  Blue-shifted emission is found to the north, west and south-west of the nucleus,
and redshifted velocities are found to the north- and south-east. The velocity field is very complex suggesting multiple overlapping dynamical structures. 
The dispersion map peaks on the nucleus and has extensions to the south and to the north-east. The  peak velocity dispersion is $\sigma_{\rm v}$=66 \kms\ in the nucleus.

%integrated from 1500 to 2000 km/s (ch 9-33), cell 2 2, 3sigma: 4.7 mJy, file: 30, levs 5.6 (1,3,5,7,9) km/s colour 0 - 56 km/s, blc 232,236,trc 277, 275

\begin{figure*}[tbh]
\includegraphics[scale=0.38]{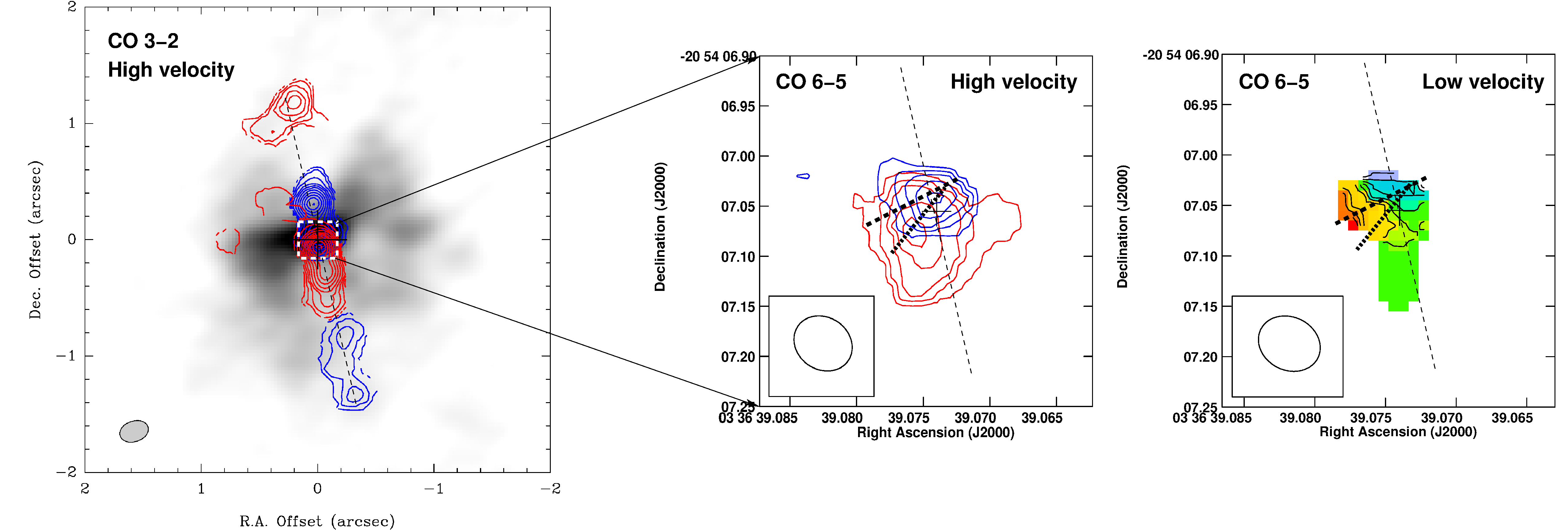}
\caption{\label{f:wings} \footnotesize Left: CO 3--2 integrated intensity image where emission close to systemic velocity (1700 - 1760 \kms)  is shown in greyscale.
The high velocity ($\pm$80 to $\pm$150 \kms) emission from the molecular jet is shown in contours (with the red and blue showing the velocity reversals). 
The dashed lines indicate the  jet axis and the inferred orientation of the nuclear disk. The CO 3--2 beam ($0.\asec 25 \times 0.\asec 18$) is shown as a grey ellipse in the
bottom left corner. (This is Fig. 2 in \citet{aalto16}). Centre: The high velocity (1500 to 1670 and 1825 to 2000 \kms)  CO 6--5 emission (see Sect.~\ref{s:hivel}). The contours
are 0.2$\times$(1,3,5,7,9) Jy \kms. The
beam is 15 times smaller (by surface) than the CO 3--2 beam. Right: The velocity field of the brightest emission (see Sect.~\ref{s:lovel})  which is close to systemic velocities. 
The colour scale is rangeing from 1690 to 1790 \kms\ and the contours start at 1690  \kms\ with steps of 10 \kms.
}
\end{figure*}

\subsubsection{High-velocity gas}
\label{s:hivel}

We integrated the high-velocity emission (1500 to 1670 and 1815 to 2000 \kms). We find that the most blue-shifted emission is located in the northern part of the nucleus.
The red-shifted emission is found in the southern part of the nucleus and in a 0.\arcsec 1 extension to the south. 
The peaks of the integrated red- and blue shifted emission are separated by 0.\arcsec 05 (5 pc) with a PA=140$^{\circ} \pm 10^{\circ}$ (see Fig.~\ref{f:wings} centre panel).

\subsubsection{Low-velocity and systemic gas}
\label{s:lovel}

The low velocity, near-systemic (1680 - 1800 \kms) gas is located in the nucleus and in 10 pc narrow gas extensions. A velocity field (moment 1) map, where only the brightest emission
(6$\sigma$) has been selected,  reveal a 100 \kms\ velocity gradient (from 1690 to 1790 \kms) along a 0.\asec 08 axis with a PA of 120$^{\circ} \pm 10^{\circ}$ (Fig.~\ref{f:wings},
right panel). The orientation of the velocity shift is roughly consistent with that of the CO 3--2 disk-like feature \citep{aalto16}). 

In Fig.~\ref{f:wings} there is also a systemic-velocity feature extending south by 0.\arcsec 1.  A position-velocity (PV) diagram shows this structure as prominent, collimated emission
along the north-south axis (Fig.~\ref{f:pvns} discussed in Sec.~\ref{s:jet}).   There is also a narrow, somewhat redshifted 
(1780-1800 \kms), emission component streching out to the north-east (Fig.~\ref{f:Keplerian} discussed in Sec.~\ref{s:hivel}).
%There is an additional lump of emission 20 pc to the north-west. 

%

\subsection{Continuum and HCN 8--7}

We merged all line-free channels in our observations, but we do not detect any 690~GHz continuum with an upper limit of 2 mJy (1$\sigma$). 
%In our previous ALMA band 7 data we find only very faint 345 GHz continuum consisting of a compact component and some extended emission (1.3$\pm$0.1 mJy beam$^{-1}$ peak and
%2.2$\pm$0.3 mJy integrated). The 345 GHz continuum brightness temperature is 0.3 K in a 20 pc beam. For the 690~GHz the $T_{\rm B}$(dust) is $<$ 2 K in the 5 pc beam.
No HCN 8--7 emission was detected with limits to the integrated intensity of 2 Jy \kms.

%Note for tapered beamsizes 0.14,0.25,0.8 the onesigma limit is 3.5, 4.6, 13.1 mJy.

%------------------------------------- Discussion ----------------------------------------------------------------------------------
%-----------------------------------------------------------------------------------------------------------------------------------

\section{Nuclear gas and dust properties}

\label{s:nucleus}

\subsection{Gas excitation}
\label{s:ex}

We convolved our CO 6--5 map to the same resolution (25 pc) as that of the CO 3--2 ALMA data of \citet{aalto15b}.  The average $T_{\rm B}$ for CO 3--2 is 34 K , and for CO 6--5 it is $40$ K
resulting in a CO 6--5/3--2 intensity ratio of ${\cal R}$ of 1.2.  Such a high value of ${\cal R}$ implies gas densities $n \gapprox 10^4$ $\cmmd$, kinetic temperatures $T_{\rm kin} > 100$ K, 
and low to moderate ($\tau \lapprox 1$) line opacities. These properties are similar to those found in the nucleus of the Seyfert galaxy NGC~1068 on spatial scales of 35 pc \citep{garcia14,viti14}. 

Using the RADEX \citep{vandertak07} non-LTE radiative transport model, we find that CO column densities of $N({\rm CO})=3 \times 10^{18}$ $\cmmt$ (for $\Delta V$=150 \kms) and line optical
depths near unity, can satisfy ${\cal R}$ and the observed line $T_{\rm B}$\footnote{This $N({\rm CO})$ has an error of a factor of a few. For higher $N({\rm CO})$, ${\cal R}$ quickly drops below unity. For lower $N({\rm CO})$ it becomes increasingly difficult to find solutions with sufficient line brightness temperature.}. For gas number densities $n \approx 10^4$ $\cmmd$, gas kinetic temperatures are high, $T_{\rm kin} > 300$ K and for higher densities $n \approx 10^5$ $\cmmd$ temperatures can be lower, $T_{\rm kin} >100$ K. 

For typical CO abundances of $10^{-5} - 10^{-4}$, the $N({\rm CO})$ implies $H_2$ column densities of $N$(H$_2$)=$3 \times 10^{22} - 3 \times 10^{23}$ $\cmmt$. This 
is about one order of magnitude lower than that derived from the CO conversion factor (see \citet{aalto16}). If there is a steep nuclear temperature gradient, the value of 
${\cal R}$ may be $>>$1 in the inner 5 pc to $<$1 at a resolution of 25 pc. This could potentially allow for higher values of $N$(H$_2$), but the weak 345~GHz (860~$\mu$m) continuum in the nucleus
found by \citet{aalto16} places constraints on $N$(H$_2$).  In the 345~GHz ALMA continuum map we find a total flux of 2.2 mJy in a structure with
a FWHM size of 0.\asec 25 $\times$ 0.\asec 09. The peak flux is 1.3 mJy which corresponds to $T_{\rm B}$(345 GHz)=0.3 K.  A lower limit to $T_{\rm d}$ is 34 K which results in an upper limit to column densities of  $10^{23}$ $\cmmt$ (for a gas-to-dust ratio of 100 and following the prescription in \citet{keene82}). If temperatures are indeed 100 K the gas column densities drop to $3 \times 10^{22}$ $\cmmt$. 

%Half of the flux is distributed in an even more low surface-brightness 0.\asec 8 structure \citep{aalto16}. 
%If there is a steep radial temperature gradient, we may have dust tempertures in this region around $T_{\rm d}$=30-50 K. 
%However, the observed flux is consistent with an $N$(H$_2$) of $10^{22}$ $\cmmt$. Even at $T_{\rm d}$=10 K, column densities are well below $10^{23}$ $\cmmt$. 
%Finally, we have the 1.3 mJy 345 GHz continuum observed in the 0.\asec 25 beam. 

%------------------------------------------------------------------------

\subsection{How obscured is NGC~1377?}
\label{s:obscured}

NGC~1377 is a galaxy with a deep silicate absorption implying a deeply enshrouded nucleus \citep{spoon07} and it has been suggested to belong to a group of galaxies with 
Compact Obscured Nuclei (CONs). These objects have dust continuum emission that is optically thick down to mm wavelengths (e.g.  the LIRG NGC4418 \citet{sakamoto13, costagliola13}). CONs have opaque structures on scales of tens of pc with $N$(H$_2$) in excess of $10^{25}$ $\cmmt$.  CONs likely host young dust enshrouded nuclear activity in the form of compact starbursts and/or accreting SMBHs \citep{aalto15b}. NGC~1377 has also been proposed to harbour young nuclear activity at least partially powered by an accreting SMBH \citep{aalto12b,aalto16, costagliola16}. Its lack of X-ray emission has been suggested to be caused by a large, Compton Thick (CT), column of obscuring dust and gas with $N$(H$_2$) $>10^{24}$ $\cmmt$ \citep{costagliola16}.

Compact 690~GHz continuum emission is, however, not detected in the nucleus of NGC~1377 with a limit of $T_{\rm B}<2$ K  (2 mJy). This is consistent with the faint
345~GHz and 230~GHz continuum found with ALMA and SMA \citep{aalto12b,aalto16} on larger scales (see below).  For $T_{\rm B}(690 {\rm GHz})<2$ K and a dust temperature 
$T_{\rm d} \gapprox$100 K, we can estimate an upper limit to the optical depth $\tau(690 {\rm GHz})$ of 0.02. From \citet{keene82},  the corresponding $N$(H+H$_2$) is $\lapprox2.4 \times 10^{23}$ $\cmmt$ in the 5 pc ALMA beam. This is consistent with $N$(H+H$_2$) derived from the 345 GHz continuum emission in a 0.\asec 25 aperture (see Sec.~\ref{s:ex}).
%The resulting average gas density is $n \lapprox 10^4$ $\cmmd$.
The limit to $N$(H+H$_2$) is also similar to the hydrogen column density estimated from the silicate absorption feature \citep{roussel06,lahuis07}. 
This is a substantial column density, but it does not suggest that the nucleus of NGC~1377 is CT. Not even soft (0.5-2 keV) X-rays become strongly absorbed at these
column densities \citep{treister11}. However, CT obscuration may occur on smaller or larger scales. 
Lower resolution {\it Herschel} observations, for example, find a 690~GHz flux of 200 mJy \citep{dale14} in contrast to the 2 mJy nondetection in our 5 pc ALMA beam. The 690~GHz flux detected by
{\it Herschel} must therefore have an extended distribution with lower surface brightness. In the following two sections we will discuss the possibility that either foreground obscuration or a small, subparsec strcuture may provide line-of-sight CT obscuration.

\subsubsection{Foreground obscuration}
\label{s:foreground}

\begin{figure}[tbh]
\includegraphics[scale=0.30]{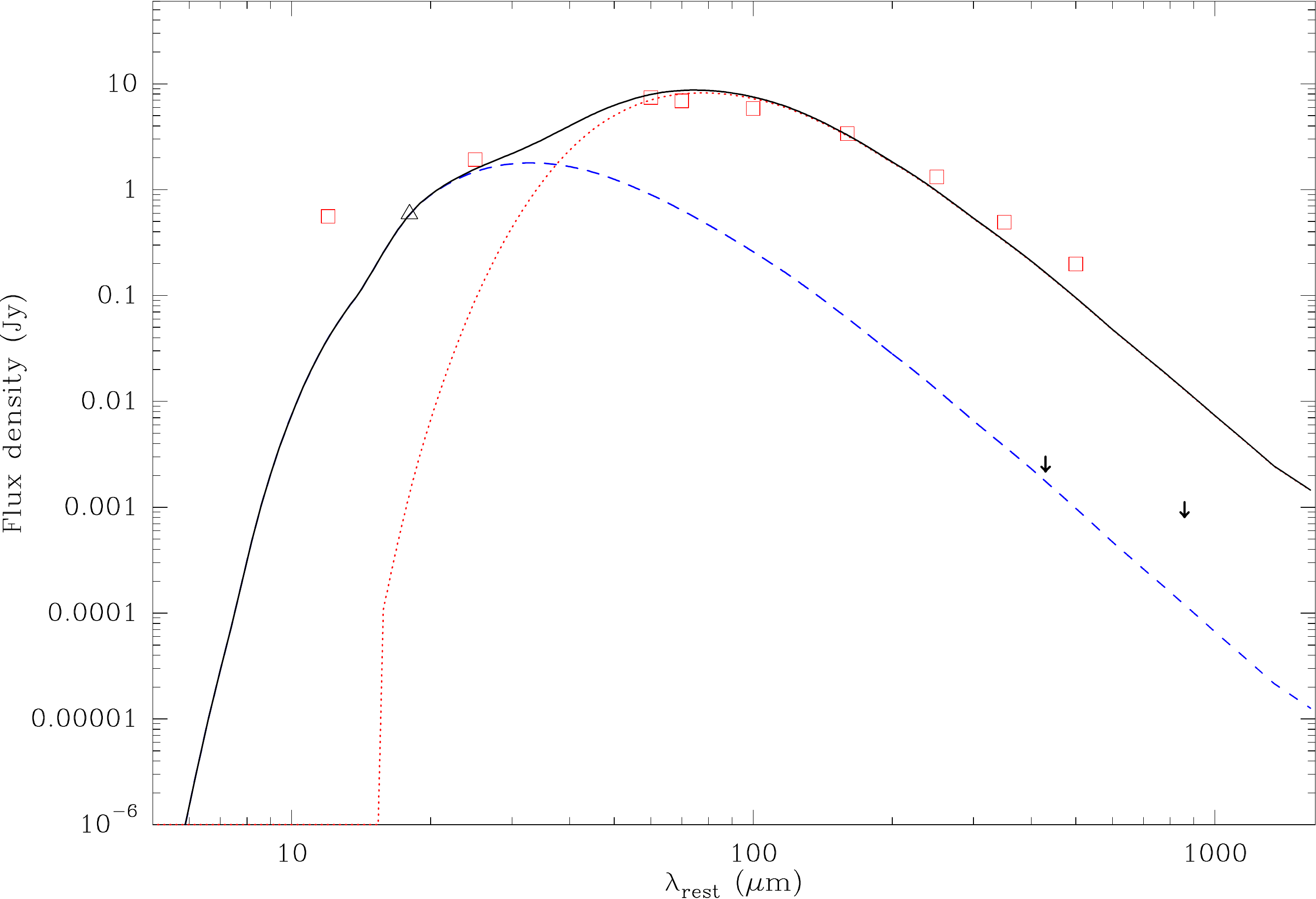}
\caption{\label{f:models}  \footnotesize An example of a two-component model for the dust SED.  It consists of a hot compact (0.\asec 14) component with $T_{\rm d}$=100 K and an extended (1.\asec 5) component with $T_{\rm d}$=40 K. For a gas-to-dust ratio the H$_2$ column densities are $N$(H$_2$)=$3 \times 10^{22}$ and $2 \times 10^{22}$ respectively. Luminosities are $1.1 \times 109$ \lsun\ for
the compact component and  $2.2 \times ^10^9$\lsun\ for the extended one. The observed SED points are marked as red squares and were taken from http://ned.ipac.caltech.edu/forms/photo.html.
}
\end{figure}

In Fig.~\ref{f:models} we present the global IR SED of NGC~1377. We include a two-component fit to the data-points consisting of a {\it compact}
and an {\it extended} source.  We fix the source size of the compact component to 0.\asec 14 (14 pc). This is the upper limit to the high-surface brightness mid-IR (18 $\mu$m) source found by
\citet{imanishi11}. We also assign it a $T_{\rm d}$ of 100 K, since \citet{imanishi11} derive a mid-IR surface brightness of $\Sigma_{\rm mid-IR}>2.5 \times 10^{13}$ \lsun kpc$^{-2}$
implying a $T_{\rm B}(18 \, \mu{\rm m}) \gapprox$100 K. The limit to the 690~GHz flux, if we convolve our ALMA observations to a 0.\arcsec 14 aperture, is 3.5 mJy, and we use the 345~GHz
continuum flux of 1.3 mJy (in a 0.\arcsec 25 aperture \citep{aalto15b}) as an upper limit to the flux in the 0.\asec 14 aperture. The resulting compact SED is shown as the blue curve in Fig.~\ref{f:models}. 
The 12 $\mu$m point in the SED is underpredicted suggesting that $T_{\rm d}$ is $>$100 K or an additional, hotter component.

The extended source, to which we fit the IRAS FIR points and the 200 m~Jy {\it Herschel} 690~GHz continuum data point, does not have a similar size constraint. 
In general, however, we expect the low-$J$ CO emission to be at least roughly co-existent with the extended dust component and therefore we selected
a size of 1.\asec 5  \citep{aalto12b,aalto16} for the extended component\footnote{The sensitivity
of the 690~GHz ALMA observations presented here declines with increasing beam-size and in a 1.\asec 8 beam there are only very few baselines left and the sensitivity is
effectively 50 mJy. There is therefore nothing in our observations that contradict that the {\it Herschel} flux can be distributed on the scales inferred from the CO observations.}.
We find that the compact component has $N$(H+H$_2$) in the range of $10^{22} - 10^{23}$ $\cmmt$
and to the extended component we can fit values of a few times $10^{22}$ $\cmmt$ for dust temperatures around 40 K. These combined components will not provide
CT obscuration, but we require sensitive, lower resolution ALMA observations to search for the potential presence of colder large-scale emission. The 345 GHz ALMA observations
hint at the presence of a disk-like 0.\asec 8 continuum feature with very low surface brightness. This structure should be explored in future observations.

%It is probably possible to come up with a model  that lines up  $\theta$=0.\asec 14, 0.\asec 25, 0.\asec 8 and 1.\asec 5 components to finally add up to an $N$(H$_2$) of $10^{24}$ $\cmmt$.
%This would however require the 0.\asec 25, 0.\asec 8 to be very cold. 
 
%and also consistent with the $N$(H$_2$) we derived (on 25 pc scales) in Sect.~\ref{s:ex} above.

\subsubsection{Is there a subparsec Compton thick core?}
\label{s:compton}
Alternatively, a CT source may be significantly smaller than our ALMA 5 pc beam. A simple model with $N$(H$_2$)=$1 \times 10^{25}$ $\cmmt$ and constrained to have
an upper limit to its 690~GHz continuum flux of 2 mJy gives corresponding maximum
sizes of an opaque dust structure in the NGC~1377 nucleus. 
%Model dust Spectral Energy Distributions (SEDs) are presented in Fig.~\ref{f:models} for four different dust temperatures
%100, 300, 500 and 700 K. 
We can model the 690~GHz flux from such an opaque structure through $S(690~GHz)=10^{-26} \, B_{\nu}(T) \, \pi \, r^2/D^2 \, (1-exp({-\tau}))$ (where $D$ is the distance to NGC~1377,  $r$ is the radius of the dust structure, $B_{\nu}(T)$ is the Planck function). Thus, for an upper limit on S(690), we have $r=\sqrt(0.002/1e26/B_{\nu}(T)/\pi D^2/(1-exp(-\tau)))$. Maximum sizes for $r$ range from 0.7 to 0.2 for dust temperatures ranging from 100-700 K. The dust sublimation radius is  0.02 - 0.05 pc from an AGN of luminosity $10^{43}$ erg/s \citep{netzer15}. An obscuring screen as small as $r$=0.2 pc may therefore be possible and would not be detected with the resolution and sensitivity of the ALMA obsrvations presented here.
%The presence of such a small high column density structure would imply that the powerful molcular outflow of NGC~1377 has expelled molecular gas from within a radius of 5-50 pc, while leaving an opaque %structure in the core. 

%Fig.~\ref{f:models} shows that for $T_{\rm dust}$ of 100 K the model SED underpredicts the observed flux (red squares) on all wavelengths while for $T_{\rm dust} %\gapprox$300 K the
%model dust core could produce the observed flux at 10-20 $\mu$m and is consistent with the ALMA 690~GHz continuum nondetection. 

\subsection{Origin of the FIR luminosity}

The detection of faint radio emission with a synchrotron spectrum by \citet{costagliola16} suggests that NGC~1377 is not powered by a buried nascent starburst.
This is consistent with the discovery of a molecular jet in NGC~1377 which suggests that its nucleus is powered by a single accreting object, an AGN.
In addition, based on their IR AGN-Starburst diagnostic diagrams, \citet{dale14} suggest that NGC~1377 is located away from "actively star forming" and into the region with
a significant AGN contribution (near the 75\% contribution line).

\smallskip
\noindent
We have previously estimated the mass of the buried SMBH to $\approx$ $1.5 \times 10^6$ \msun\ \citep{aalto12b}. The Eddington luminosity of a 10$^6$ \msun\ SMBH is about
$5 \times 10^{10}$ \lsun. If all the mid-IR luminosity ($1 \times 9$ \lsun\ to $4 \times 9$ \lsun) is due to an accreting SMBH,  we find it is growing at a rate of 1\% - 10\% Eddington which
would place it in the quasar-mode of accretion\footnote{Note that this discussion assumes that the IR emission is absorbed and reradiated uv and X-ray emission from an efficiently accreting SMBH. An alternative possibility is an AGN in low-accretion radio-mode. It may generate a radio jet which is shock heating gas and dust in the nucleus. This scenario would be unusual but deserves future  consideration - for example through searching for hard X-ray emission associated with hard-state low accretion mode.}. Alternatively, the mid-IR emission may be originating from an extremely radio-quiet nuclear starburst as suggested by \citet{roussel06}. The molecular jet would then be powered by an SMBH accreting at a substantially lower Eddington rate.

In our models above, the FIR emission cannot be emerging from the same compact structure as the mid-IR. We inferred a source size of 1.\asec 5 to fit with the CO emission, but we stress
that we do not know the extent of the FIR emission. We require a lower resolution and higher sensitivity image to locate the missing 690~GHz emission and to solve the mystery of what
processes generate the FIR emission of NGC~1377. However, it is reasonable to assume that at least part of the FIR and CO ($J$=2--1, 3--2) emission coexist and we note that most of the CO 2--1 and 3--2 emission of NGC~1377 appears to be associated with a molecular jet/outflow \citep{aalto12b,aalto16}. The lack of star formation indicators imply that a substantial part of the FIR emission is not associated with dust heated by massive stars. An alternative is that the dust is heated by mechanical interactions in the outflow, and/or by photons escaping from the AGN along the poles. The old stellar component may also contribue to the heating of the dust and gas.

%----------------------------------------

\section{Nuclear dynamics}
\label{s:hivel}

The observations were primarily designed to locate a (purported) high-surface brightness nuclear continuum source with associated compact CO 6--5 emission. In addition, observations of 
Galactic disk-outflow objects show that the CO 6--5 emission is often associated with current shock regions (while lower-$J$ emission such as CO 3--2 more reflects conditions in the larger-scale environment)
\citep{yildiz15}. Thus, our CO 6--5 observations may not give us a complete picture of the dynamics in the nuclear region. However, they provide important information on the very inner kinematics and how the activity is interacting with its surroundings.
The PA of the integrated CO 6--5 emission  20$^{\circ} \pm15 ^{\circ}$ implies that it is largely aligned with the jet/outflow and not dominated by a rotating torus. CO 6--5 emission from the
base of the jet/outflow shows that the gas is dense ($>10^4$ $\cmmd$) and high brightness temperatures reveal that it is also warm. Possible heating mechanisms include shocks in the outflowing gas
or from instabilities, and/or heating by radiation from the AGN. 

The gas at  the highest velocities (that we are sensitive to) is found very close ($r$=2-3 pc) to the nucleus. The velocity shift is 145 \kms, which is similar to the shift of 150 \kms\
found previously for the lower resolution CO 3--2 data. However, there the maximum velocity occured further from the nucleus (0.\arcsec 25=25 pc) along a collimated jet-structure \citep{aalto16}. The velocity reversals along the CO 3--2 jet (see Fig.~\ref{f:wings}, left panel) are suggested to be caused by jet precession \citep{aalto16}. Below we discuss if the
CO 6--5 high velocity gas is associated with the rotation of a nuclear disk or with the jet/outflow.

\subsection{The high velocity gas}

\begin{figure}[tbh]
%\centering
\includegraphics[scale=0.31]{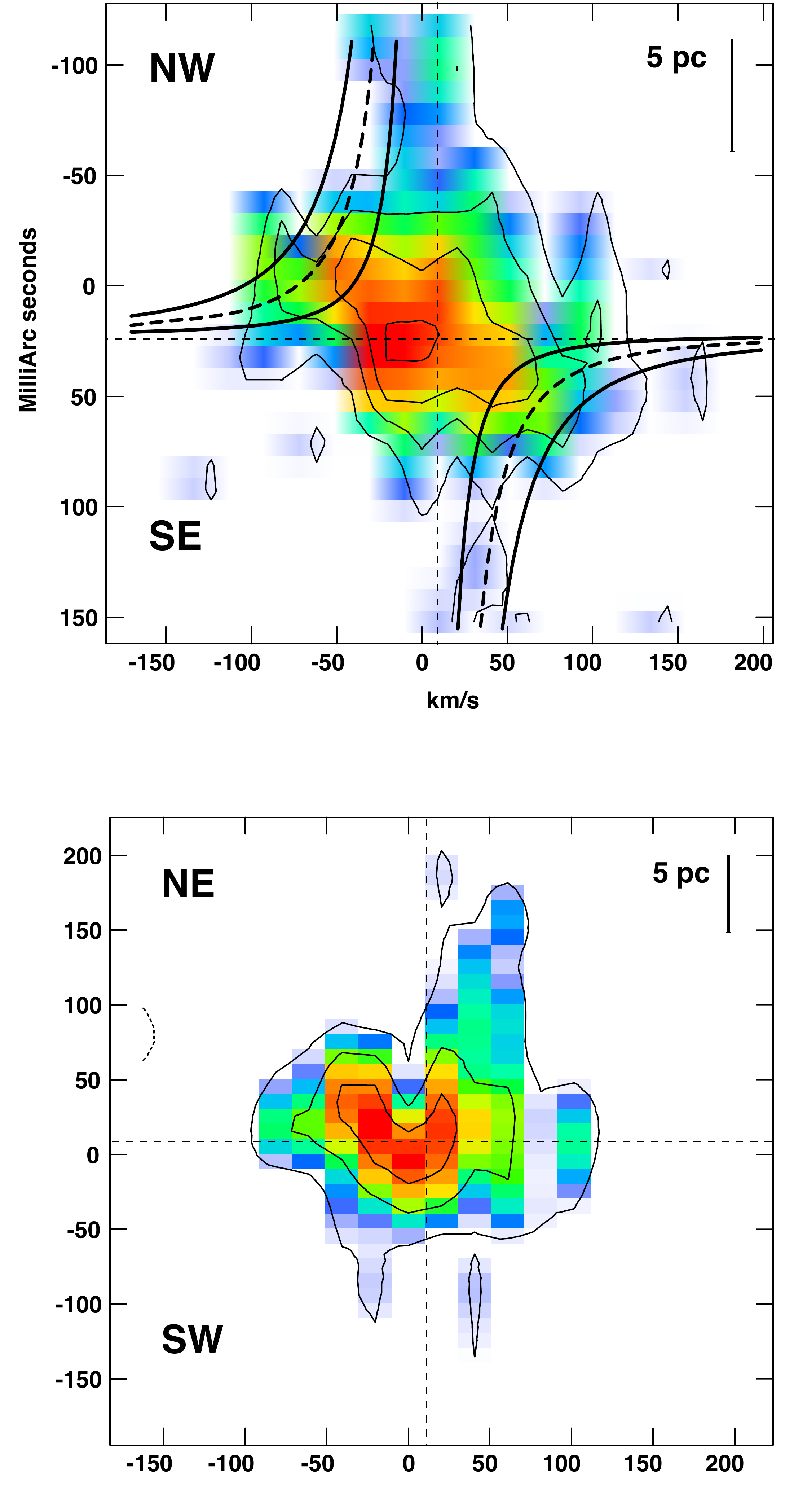}
\caption{\label{f:Keplerian}  \footnotesize CO 6--5 PV diagrams. Top: PA=135$^{\circ}$ close to the major axis of the nuclear disk. The curves show the Keplerian rotation curves of a: $10^6$, $3 \times 10^6$ and $6 \times 10^6$ \msun\ SMBH. Bottom: Cut perpendicular (PA=45$^{\circ}$) to the nuclear disk. 
The contour levels are 25$\times$ (1, 2, 3)  mJy  beam$^{-1}$. Colours range from 25 to 90 mJy beam$^{-1}$. A velocity of 0 \kms\ corresponds to 1743 \kms.
}
\end{figure}

\noindent
{\it Disk} \, If the velocity shift is due to rotation in a disk, the projected rotational velocity is 73 \kms.
The peaks are separated by 5 pc and, assuming an edge-on disk, the dynamical mass\footnote{A simple estimate of the dynamical mass is  $M_{\rm dyn}$=RV$\mathstrut{^{2}_{{\rm rot}}}$/G, where V$_{\rm rot}$ is the rotation speed.} is $M_{\rm dyn}=3 \times 10^6$ \msun\ inside $r$=2.5 pc. This is close to the SMBH mass of $1.5 \times 10^6$ \msun\ inferred from the  $M$-$\sigma$ relation \citep{aalto12b}. The PA between the peaks of 140$^{\circ} \pm 10$ is shifted by 20$^{\circ}$ from that of the lower velocity rotating structure, but it is within the error bars of both PA. Thus, it is possible that the high velocity peaks (at least partially) may stem from a Keplerian disk. 
In Fig.~\ref{f:Keplerian}  we show the Keplerian tracks of three different compact masses overlayed on the major axis PV diagram of the high velocity emission. Comparing the tracks to the data we see
that the expected high velocity cusp in the center is missing. Deeper observations are required to determine if this is a sensitivity issue or if there is no nuclear gas with velocities
$>$100 \kms. There is also CO 6--5 emission at forbidden velocities indicating non-circular motions in the form of turbulence, instabilities and/or outflowing gas \citep{garcia16}. 
Within the errors, the data allows for either a single SMBH or a combination of a SMBH and a compact nuclear stellar cluster (see e.g. cluster rotation curves in \citet{stone15}).

We can obtain estimates to the disk gas inside 5 pc by adopting the limit to $N$(H$_2$) derived in Sec.\ref{s:obscured} and assume that it fills the beam, which leads to
$M_{\rm gas} \lapprox 10^5$ \msun.  However, since there may also be emission from the outflow in the inner beam, a limit to 
$M_{\rm gas}$ can also be obtained through inferring an edge-on disk with thickness $h$=$r/2$ and using the limit to $N$(H$_2$) then $M_{\rm gas}\lapprox 2 \times 10^4$ \msun.
In any case, the gas mass is significantly lower than the dynamical mass ($<$3\% ), which again suggests similar properties to those of the non self-gravitating gas torus of NGC~1068 \citep{garcia16}.

\medskip
\noindent 
{\it Jet/outflow} \, The previously found molecular jet has a symmetry axis of $\approx11^{\circ}$ and was suggested to precess with an angle $\theta$=10$^{\circ}$ - 25$^{\circ}$.  
The simple precessing jet model proposed in \citet{aalto16} predicts that velocities, after peaking at a distance $\pm$25 pc, decrease closer to the nucleus. In the CO 6--5 data there are no
strong signatures ($>$3$\sigma$) of high velocity gas at $\pm$0.\asec 25.  This may be a senstivity issue and/or that the gas temperature and density is only high enough to excite the 
CO 3--2 line 25 pc along the jet - not CO 6--5.  The PA of 140$^{\circ} \pm 10$ of the nuclear CO 6--5 high-velocity gas is also inconsistent with it being associated with gas along the jet.
However, luminous CO 6--5 emission at velocities closer to systemic is detected along the jet (see discussion in Sec~\ref{s:jet}).

We tentatively conclude that the most likely origin of the nuclear CO 6--5 high velocity gas is in a rotating, inclined disk. However, we cannot exclude that the jet behaviour has changed close
to the nucleus or that it has turned off. We may be resolving the jet or the high velocity gas is emerging from another outflow structure.

\begin{figure}[tbh]
\includegraphics[scale=0.35]{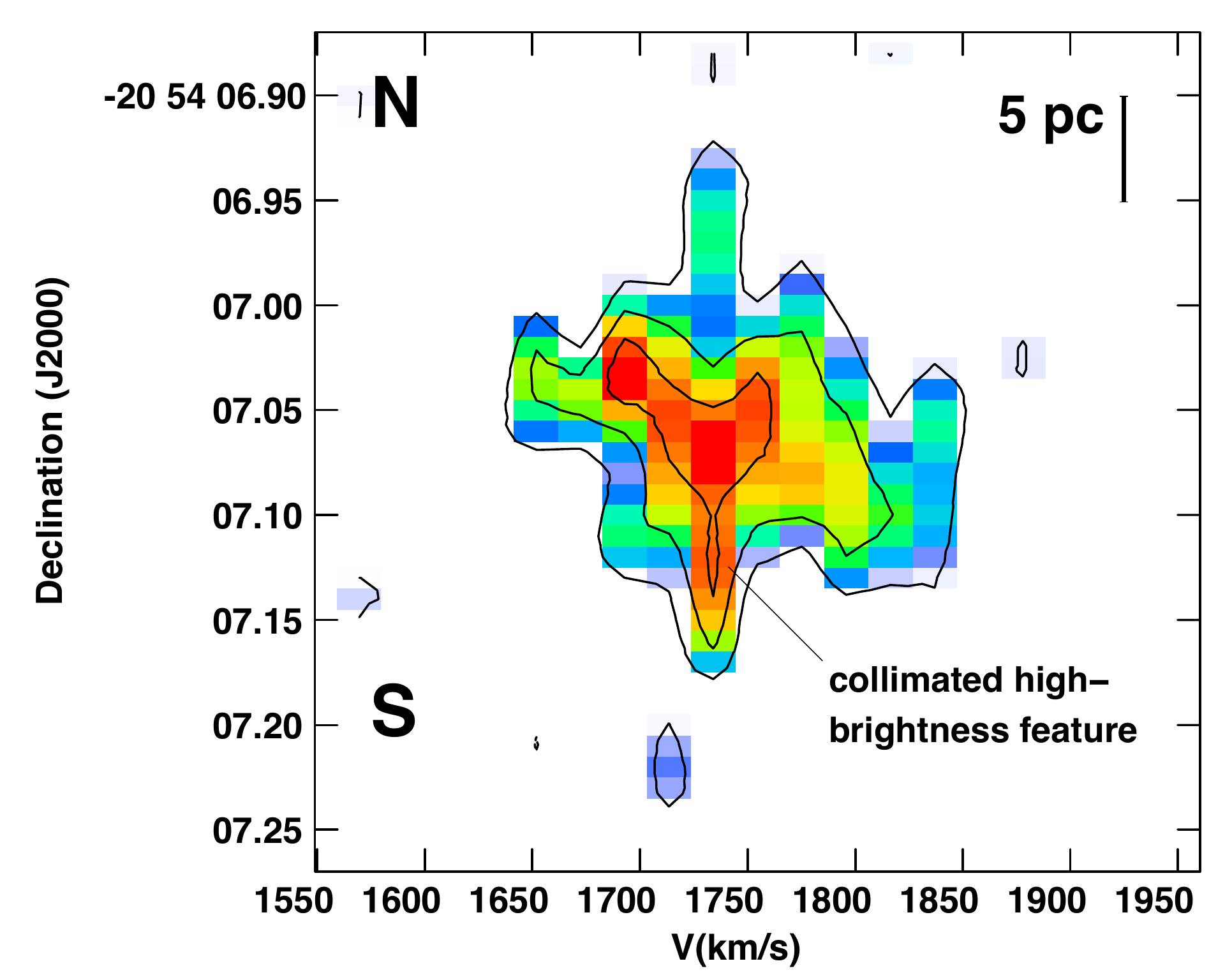}
\caption{\label{f:pvns} \footnotesize Position-velocity (PV) diagram along the north-south axis. The contour levels are 25$\times$ (1, 2, 3)  mJy  beam$^{-1}$. Colours range
from 25 to 90 mJy beam$^{-1}$. 
}
\end{figure}

%----------

\subsection{Jet dynamics}
\label{s:jet}

A component that would fit the jet-precession model is the bright, narrow feature emerging from the nucleus in the north-south direction (Fig.~\ref{f:pvns}). It is found at systemic velocities,
and would be consistent with a scenario where the jet has precessed away from its symmetry axis and is now at its maximum angle away from the symmetry axis.
In this case, the jet precession angle would be close to $\theta$=10$^{\circ}$. The high CO 6--5 brightness
temperature of (at least the southern) collimated feature may indicate shock-heating by the jet-interaction.  Note that if jet precession is caused by the warping of the nuclear disk
(see discussion in \citet{aalto16}) then the inner disk should have a PA close to 90$^{\circ}$ which appears to be inconsistent with the PA of the nuclear disk found above. However, 
the precession may be governed by the accretion disk which would be much smaller than 5 pc.
%If the near-nucleus jet-orientation is north-south it is not perpendicular to the putative nuclear disk with PA=120$^{\circ}$ - 140$^{\circ}$. The jet orientation may not be perpendicular to the 
%molecular disk - or the very nuclear disk may be warped on scales too small for us to see. 

There are also other (somewhat fainter) narrow jet-like features in the map. One is a structure emerging to the north-east at PA=45$^{\circ}$ and which may have a 
counterpart to the south-west. We can se it as a redshifted component extending 20 pc along the north-eastern part of the minor axis (right panel in  Fig.~\ref{f:Keplerian}).
It is not clear if this is a separate component - or if all (seemingly) collimated features are part of the same outflow.  
In our previous CO 3--2 observations, we did not resolve the high velocity jet in our 0.\asec 25 beam. Even though a radio jet may have a very narrow width and could be unresolved
also by our CO 6--5 5 pc beam, it would interact with its surroundings resulting in a wider structure. 
% If it is static, and does not precess, it will mostly impact material
%in its path that it runs into, or through turbulent entrainment of gas along the jet. A wider angle or precessing jet may affect its environment in greater deal. 

\subsubsection{A disk-wind?}

It is also possible that the jet formation is strongly linked to the molecular disk in the form of a disk-wind (see \citet{gallimore16} for a discussion of a potential disk-wind scenario
for NGC~1068). In this case we may expect to see remnants of the disk rotation in the outflowing gas as a
velocity gradient across the jet \citep[e.g.][]{smith07,launhardt09}. Jet-rotation is an efficient way of removing angular momentum and solving the problem of extracting the angular momentum of
the circumnuclear gas. Jet/outflow rotation may also be driven by MHD shocks in helical magnetic fields \citep[e.g.][]{fendt11}. Jet rotation structures dissipate quickly and are expected to be found
close to the nucleus. 

There is a velocity shift across the jet to the north and south of about 150 \kms, but it is not clear if this is indeed due to jet rotation or to overlapping, separate
dynamical components stemming from other processes. There may also be contamination from the disk rotation near the nucleus. In addition, a very thin radio-jet pushing itself through
a dense medium may result in gas being accelerated in opposite directions, perpendicular to the jet (see e.g. discussion in \citet{dasyra15}). To fully explore the link between the disk and the
jet/outflow high-resolution observations at multiple frequencies are required.

%----------

%Continuum models:
%100K/0.66pc ger 2.7e7 L_sun
%300K/0.36pc ger 6.4e8 L_sun
%500K/0.28pc ger 2.5e9 L_sun
%700K/0.23pc ger 4.9e9 L_sun

%\subsection{IR clues}

\section{Conclusions}

We have used ALMA to image the CO 6--5 emission, and put limits on
the 690~GHz continuum in the nucleus of the extremely radio-quiet lenticular galaxy NGC~1377 with a resolution of
6 $\times$ 5 pc. We find luminous, compact (10 $\times$ 7 pc)  CO 6--5 emission from hot ($T_{\rm kin} \gapprox 100$ K) molecular 
gas.  The CO 6--5 integrated intensity is aligned with the previously discovered
jet/outflow of NGC~1377 and is also tracing the dense, hot gas at the base of the outflow. Collimated, 10 pc long, high brightness features extend from the nucleus and are likely
associated with the jet. The velocity structure is complex and a gradient across the jet may be a signature of rotation, alternatively the gradient is caused by separate, overlapping dynamical
components.

High velocity gas ($\Delta v \pm145$ \kms) is detected inside $r<$ 2-3 pc and we suggest that it is emerging from an inclined rotating disk or torus
of PA 140$^{\circ} \pm 20^{\circ}$ and a dynamical mass of $3 \times 10^6$ \msun. This mass is consistent with that of a supermassive black hole (SMBH), as inferred from
the $M-\sigma$ relation. The gas mass of the proposed disk/torus constitutes $<$3\% of the nuclear dynamical mass.
In contrast to the intense CO 6--5 line emission, we do not detect dust continuum with an upper limit of $S$(690~GHz)$\lapprox$2mJy. The corresponding 5 pc H$_2$ column
density is estimated to $N$(H$_2$)$<3 \times 10^{23}$ $\cmmt$ which is inconsistent with a Compton Thick (CT) source (with $N$(H$_2$)$>10^{24}$ $\cmmt$). CT obscuration may instead occur
in a much smaller, subparsec, structure undetected by ALMA.  CT obscuration stemming from larger scale structures seems inconsistent with current data.
From SED fitting we suggest that half of the IR emission of NGC~1377 is nuclear and the rest (mostly the far-infrared (FIR)) is emerging from larger scales.  The extreme radio quietness, and
the lack of other star formation tracers, raise questions on the origin of the FIR emission. We discuss the possibility that it is arising from the dissipation of shocks in the molecular jet/outflow or
from irradiation by the nuclear source along the poles. If the FIR emission is associated with star formation in a disk, it is for some reason not producing radio emission.

\medskip

\begin{acknowledgements}
This paper makes use of the following ALMA data:
    ADS/JAO.ALMA\#2012.1.00900.S. ALMA is a partnership of ESO (representing
    its Member States), NSF (USA) and NINS (Japan), together with NRC
    (Canada) and NSC and ASIAA (Taiwan), in cooperation with the Republic of
    Chile. The Joint ALMA Observatory is operated by ESO, AUI/NRAO and
NAOJ.
We thank the Nordic ALMA ARC node for excellent support. SA acknowledges
support from the Swedish National Science Council grant 621-2011-4143. 
F.C. acknowledges support from Swedish National Science Council
grant 637-2013-7261 
KS was supported by grant MOST 102-2119-M-001-011-MY3
SGB thanks support from Spanish grant AYA2012-32295. 

\end{acknowledgements}

\bibliographystyle{aa}
\bibliography{n1377_ALMA_aalto_2}

\end{document}